# Rietveld refinement of ZrSiO$_4$: application of a phenomenological model of anisotropic peak width

A. Sarkar[*], P. Mukherjee, P. Barat

Variable Energy Cyclotron Centre
1/AF Bidhan Nagar, Kolkata 700064, India

[*] Contact author; e-mail: apu@veccal.ernet.in

**Keywords**: Rietveld refinement, anisotropic broadening, microstructure

**Abstract.** The anisotropic broadening of ZrSiO$_4$ sample is modelled using the Stephens's phenomenological model for anisotropic line broadening and the three dimensional strain distribution in the sample is plotted. The microstructural parameters like domain size and dislocation density are estimated using the variance method.

Introduction

Line broadening is a well known feature of the diffraction profiles from polycrystalline samples. The phenomenological line-broadening theory of plastically deformed metals and alloys was developed more than 50 years ago [1]. It identifies that besides the instrumental contribution there are two main types of broadening: the size and the strain components. The former one depends on the finite size of the coherent diffraction domains and the latter is caused by any lattice imperfection (point, line or plane defects). There are many approaches of line profile analysis to extract the size and strain information from the line broadening.
  The Full Width at Half Maxima (FWHM) and the integral breadth (area of the peak divided by the maximum intensity) are often considered as a measure of the broadening of the diffraction peaks. If the FWHM of the reflections of a line profile increases with the diffraction angle the broadening is called isotropic otherwise it is termed as anisotropic peak broadening.

This anisotropic line shape broadening is frequently observed in powder diffraction pattern and creates a serious difficulty for the Rietveld analysis.

The anisotropic peak broadening may arise due to variety of reasons like anisotropic size broadening, stacking fault or anisotropic strain broadening. In simple case the strain broadening is isotropic and the FWHM is proportional to *tan*θ i.e.

$$\Gamma_{2\theta} = X \tan\theta \qquad (1)$$

This relation does not hold good for anisotropic strain broadening. Anisotropic strain distribution has been introduced by several authors [2] to model the anisotropic peak broadening.

## Stephens model

Recently P. W. Stephens [3] proposed a phenomenological model of anisotropic broadening in powder diffraction considering the distribution of lattice metric parameters within the sample. In this model each crystallite is regarded as having its own lattice parameters, with multidimensional distribution throughout the powder sample. The width of each reflection can be expressed in terms of moments of this distribution, which leads naturally to parameters that can be varied to achieve optimal fits. Further description of the model can be found in the Stephens paper [3].

Let $d^*_{hkl}$ be the inverse of the $d$ spacing of the (*hkl*) reflection. Then $d^{*2}$ is bilinear in the Miller indices and so can be expanded in terms related to the covariances of the distribution of the lattice metrics. This leads to an expression in which the variance of $d^{*2}$ is a sum of 15 different combinations of Miller indices in the fourth order. Imposing the symmetry of the tetragonal lattice reduces the number of independent terms to the following four:

$$S^2 = S_{400}(h^4 + k^4) + S_{004}l^4 + 3.S_{202}(h^2l^2 + k^2l^2) + 3S_{220}h^2k^2 \qquad (2)$$

The anisotropic strain contribution to the angular width in 2θ of the reflection is given by

$$\delta 2\theta = (360/\pi)(\delta d/d)\tan\theta, \qquad (3)$$

where $\delta d/d = \pi(S^2)^{1/2}/18000 d^*_{hkl} \qquad (4)$

The Rietveld refinement package GSAS [4] has implemented Stephens's model to account strong anisotropy in the half widths of reflec-tions. Using this package the three dimensional strain field within the material can be reconstructed.

## Experiment and data analysis

Zircon ($ZrSiO_4$) is one of the technologically important oxide ceramic materials known for its high refractoriness and chemical stability. It shows excellent thermal shock resistance as a result of its very low thermal expansion coefficient ($5.3 \times 10^{-6}$ $K^{-1}$ from 25 to 1500 °C) and low heat conductivity coefficient of 6.1 W $m^{-1}$ $K^{-1}$ at 100 °C and 4.0 W $m^{-1}$ $K^{-1}$ at 1500 °C. It possesses a tetragonal crystal structure. X-ray diffraction pattern of zircon powder sample was recorded using a Philips PW3710 diffractometer with CuKα radiation (40kV, 40mA). Sample was scanned in a step-scan mode (0.02°/step) over the angular range (2θ) of 5° to 150°. X-ray diffraction data were collected for 3 sec at each step.

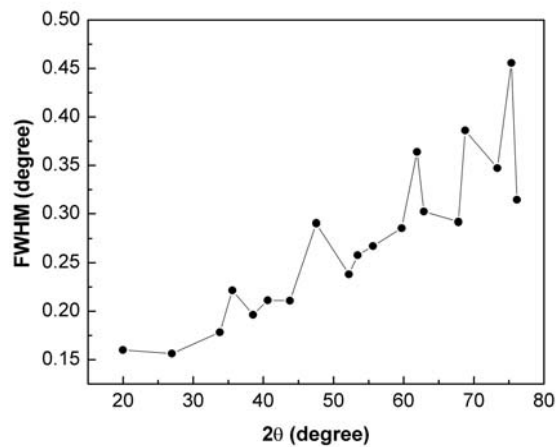

*Figure 1. Variation of FWHM with 2θ*

Figure 1 shows the variation of the FWHM of the peaks with the peak angles. The anisotropy of the width of the peaks is clear from this figure. We have used GSAS to carry out Rietveld refinement of the line profiles of the Zircon. The profiles have been fitted without and with using the Stephens model. It is found that the incorporation of Stephens's model improved the quality of the fit. A typical result of the Rietveld refinement (with Stephens's model) is shown in the Figure 2. The fitting parameter in this case was $R_{wp}$=4.65%. Without Stephen's model the fitting parameter was $R_{wp}$=8.63%. This suggests that the Stephens model fit the data very well.

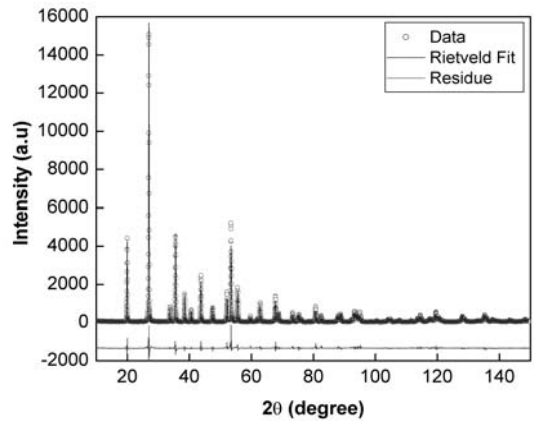

*Figure 2. Result of the Rietveld refinement*

We have used $S_{hkl}$ as the free parameters to obtain the best fit between the model and the experiment. The anisotropic broadening has both Gaussian and Lorentzian components. Therefore we have included both to get an acceptable fit of the diffraction data.

The graphical representation of the three-dimensional strain distribution can be obtained using the refined values of the $S_{hkl}$. The three-dimensional strain distribution plot of the studied Zircon sample is shown in Figure 3.

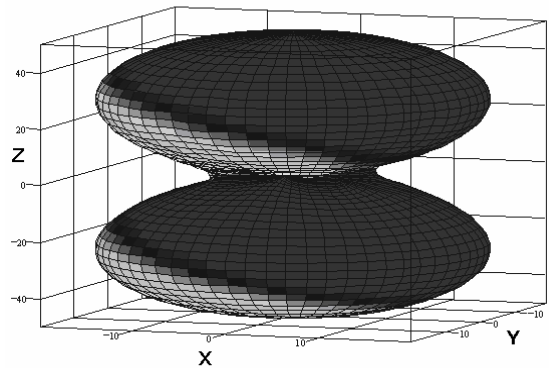

*Figure 3. Three-dimensional strain distribution of $ZrSiO_4$. The x axis is horizontal, the z axis is vertical and the y axis out of the plane of the paper. The scale is in $\delta d/d \times 10^{-6}$ strain.*

## Microstructural analysis

X-ray diffraction (XRD) is a powerful tool to characterize the microstructure of the polycrystalline samples. XRD gives the microstructure of the sample in a statistical sense. There are various methods to determine the microstructural parameters like domain size, microstrain, dislocation density from the broadened XRD peaks. We have used the *variance method* developed by Groma [5] and further modified by Borbely and Groma [6] to characterize the microstructure of the zircon sample. In the variance method one computes the k-th order restricted moment

$$M_k(q') = \frac{\int_{-q'}^{q'} q^k I(q) dq}{\int_{-\infty}^{\infty} I(q) dq} \qquad (5)$$

Here $I(q)$ is the intensity distribution as a function of $q = \frac{2}{\lambda}[\sin\theta - \sin\theta_0]$, where $\lambda$ is the wavelength of the X-ray, $\theta$ is the diffraction angle and $\theta_0$ is the Bragg angle. The second and the fourth order moments have the forms:

$$M_2(q) = \frac{1}{\pi^2 \varepsilon_F} q - \frac{L}{4\pi^2 K^2 \varepsilon_F^2} + \frac{\Lambda \langle \rho \rangle \ln(q/q_0)}{2\pi^2} \qquad (6)$$

and

$$\frac{M_4(q)}{q^2} = \frac{1}{3\pi^2 \varepsilon_F} q + \frac{\Lambda \langle \rho \rangle}{4\pi^2} + \frac{3\Lambda^2 \langle \rho^2 \rangle}{4\pi^2 q^2} \ln^2(q/q_1) \qquad (7)$$

respectively. Here, $\varepsilon_F$ is the area weighted domain size and $\langle \rho \rangle$ is the average value of the dislocation density. The details of other parameters can be found in the reference [6].

We have used the (200) peak and the variance method to estimate the microstructural parameters of the sample. Figure 4 (a) shows the (200) peak. Figure 4 (b) and (c) respectively show the variation of the second order and forth order moment of the peak. The large $q$ regions of the curves are fitted with equation (6) and (7). The values of the domain size and the dislocation density obtained from the fitting are: from $M_2(q)$, domain size=131Å and dislocation density= $5.3 \times 10^{15}$ m$^{-2}$, from $M_4(q)$, domain size=135Å and dislocation density=$4.8 \times 10^{15}$ m$^{-2}$.

## Conclusion

We have used Stephens's phenomenological model of anisotropic strain broadening to model the anisotropic broadening of the zircon sample. The Stephens model is found to very well fit the data. The microstructural parameters (domain size and dislocation density) are obtained using the variance method. The dislocation density in the sample is very high.

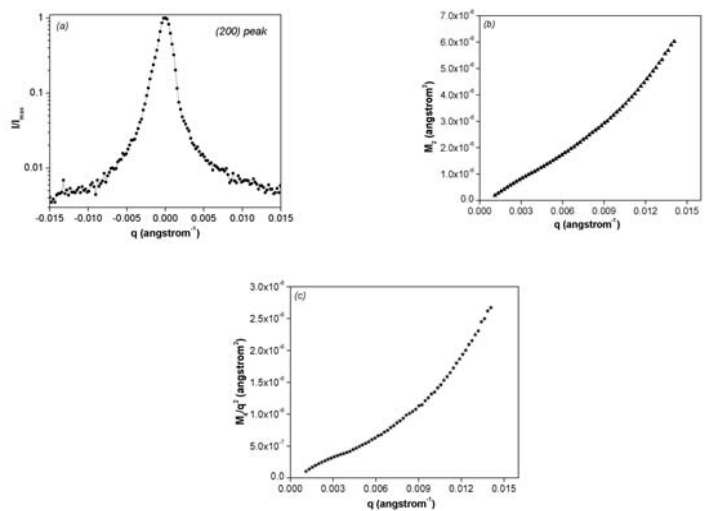

*Figure 4. (a) (200) peak (b) variation of the second moment (c) variation of the fourth moment*